# Heat and Entropy in nonextensive thermodynamics

A. K. Rajagopal
Naval research Laboratory, Washington D.C. 20375 – 5320

**ABSTRACT**
The concepts of quantity of heat and work are deduced in the non-extensive statistical mechanics context, following steps in parallel to those employed in the extensive statistical mechanics.

## 1. INTRODUCTION

It is known that the Boltzmann-Gibbs (BG) statistical mechanics fully describes the thermal equilibrium state of Hamiltonian systems[1]. In recent years, BG description is defied by a large class of non-equilibrium stationary states such as glasses. A generalization of BG formalism to accommodate such systems has been proposed by Tsallis in 1988, which will be referred here as Tsallis (T) statistical mechanics. This is accomplished by relaxing the extensivity assumption involved in the formulation of the BG entropy. For a review of this development, see [2 - 6]. This formalism closely parallels the BG theory in spite of changing the form of the entropy in having a corresponding generalized maximum entropy principle, and the Legendre transform structure required for formulating the attendant thermodynamics. At the dynamical foundation level, the ergodicity property in phase space used in BG formalism is replaced by non-ergodicity in the T formalism. One may note that while the BG theory involves exponential class of probability distributions, the T theory has power-law distributions, which are found to be quite commonly found in nature where systems are in penultimate non-equilibrium states before reaching the eventual thermal equilibrium. The purpose of this paper is to present a corresponding thermodynamic identification of terms within the T theory, even though similar work based on entirely different approach to this by Abe [7] exists. This alternate derivation is thus complimentary to Abe's work even though it parallels a different set of steps (given above for BG) in obtaining thermodynamic concepts from statistical mechanics.

In the next section we set the stage for this development by giving an account of the traditional BG approach along with a discussion of "quasi-equilibrium" based on the quantum master equation of the Lindblad type[8]. This is followed by sec.3, describing the alternate derivation mentioned above. Some comments on the corresponding quasi-equilibrium theory will also be briefly discussed. In the final sec.4, a summary of this development is given.

## 2. BOLTZMANN-GIBBS APPROACH

It is common knowledge [1] that deducing the laws of thermodynamics from statistical mechanics of many particle systems as formulated by BG depends on the identification of certain terms in the two formalisms. Let us recapitulate how this is accomplished in the standard textbooks, such as [1]. The BG statistical mechanics of a many-particle system is described in terms of the density matrix, $\rho$, given its Hamiltonian operator, H. Its total energy is defined by

$$U = Tr\rho H, \tag{1}$$

where Tr stands for the trace over the space of the many-particle system. The entropy, $S_1$, of this system is defined by the expression,

$$S_1[\rho] = -Tr\rho \ln \rho, \tag{2}$$

in units where the Boltzmann constant is taken to be unity. Using the maximum entropy principle subject to the constraints of given mean energy U defined above and the conservation of probability defined by $Tr\rho = 1$, we obtain the density matrix description of the statistical mechanical equilibrium state of the system, with

$$\rho_{eq} = Z^{-1}\exp-(\beta H), \quad Z = Tr\exp-(\beta H). \tag{3}$$

Here $\beta$ is a Lagrange multiplier, which will be identified with the inverse temperature of the system in the thermodynamics language. Z is called the partition function. One then identifies the thermodynamic quantities, the free energy F, the entropy S, and the mean energy U, associated with the system in the following way:

$$F = -\beta^{-1}\ln Z = U - \beta^{-1}S, \quad S \equiv S_1[\rho_{eq}] \tag{4}$$

This identification is in complete concordance with all the thermodynamic principles associated with the system [1]. To make connection with the thermodynamic concepts of "quantity of heat", Q, and "work", W, we follow [1], by considering a quasi-equilibrium variation of the average value of H defined in eq.(1), which clearly has two parts, one arising from change in the Hamiltonian, H, and the other arising from the concomitant change in the density matrix, $\rho$, near its equilibrium:

$$dU = TrH(d\rho) + Tr(dH)\rho. \qquad (5)$$

We then identify the quantity of heat with the first term and the work with second term:

$$dQ = TrH(d\rho), \quad dW = Tr(dH)\rho. \qquad (6)$$

By considering quasi-static processes near equilibrium, one may then capture all the thermodynamic concepts such as adiabatic and isothermal processes, Carnot cycle, etc. in the Statistical Mechanics parlance. For a discussion of this, one may refer to [9].

We now discuss briefly the quasi-static development. The most general quantum master equation which is linear, local in time, preserves the essential properties of the density matrix, namely hermiticity, traceclass (probability conservation) structure, and positive semi-definiteness is the Lindblad equation [8] (units where the Planck constant, $\hbar$ is set equal to 1):

$$\partial \rho / \partial t = -i[H, \rho] + \frac{1}{2} \sum_j k_j \left\{ 2 L_j \rho L_j^+ - L_j^+ L_j \rho - \rho L_j^+ L_j \right\} \qquad (7)$$

where the first term in the right hand side represents the unitary time evolution driven by H, a hermitian Hamiltonian operator and the second term represents dissipative evolution if all the real parameters $k_j$ are positive, accomplished by the operators $L_j$ along with their hermitian conjugates, $L_j^+$. This choice of operators is not unique and the sum over j is as yet unspecified. The unitary time evolution part may be subsumed by a transformation $\rho_d = U \rho U^+$, so that the new equation has no H-term but has modified L-operators, $\tilde{L}$. This equation has two additional features, one, it can take a pure state into a mixed state and vice versa, and two, it preserves positivity of the density matrix throughout the evolution. This last property is often violated in phenomenological master equations. Since the density matrix is hermitian, a diagonal representation for it can be chosen $\rho_d = \sum_\alpha |\alpha\rangle p(\alpha) \langle \alpha|$, where $\{|\alpha\rangle\}$ is an orthonormal set for each instant of time. $p(\alpha)$ have the same physical interpretation as the probabilities in a classical description of the system.

With the choice of the entropy functional in the von Neumann form, eq.(2), we now calculate $\partial S_1/\partial t$:

$$\partial S_1/\partial t = -Tr\,\partial\rho/\partial t \ln\rho = \sum_{\alpha,\alpha'} K_{\alpha\alpha'}\,p(\alpha')(\ln p(\alpha') - \ln p(\alpha)) \geq 0 \quad (8)$$

provided the Lindblad operators are hermitian. Here $K_{\alpha\alpha'} = \sum_i k_i |\langle\alpha|\tilde{L}_i|\alpha'\rangle|^2$. This follows after using $\ln x \geq 1 - 1/x$, $x > 0$, and upon interchanging summation indices and definitions. Thus the increase of entropy is guaranteed under the above condition. Under such a condition, one also finds that the right hand side of modified eq.(7) takes the form, $-\frac{1}{2}\sum_j k_j\left[\tilde{L}_j,\left[\tilde{L}_j,\rho_d\right]\right]$.

Another important aspect of eq.(7) concerns its stationary solution, $\rho_s$, satisfying $\partial\rho/\partial t = 0$. This will now be shown to be also the asymptotic, long time solution of eq.(7). A particular form of the stationary solution obeys the "microscopic detailed balance" relation given by

$$K_{ab}p_s(b) = K_{ba}p_s(a)\ for\ all\ pairs\ a,b. \quad (9)$$

The subscript s here denotes the stationary limit of the eq.(7) introduced here:

$$\partial\rho_{ds}/\partial t = 0 = \frac{1}{2}\sum_j k_j\left\{2\tilde{L}_j\rho_{ds}\tilde{L}_j^+ - \tilde{L}_j^+\tilde{L}_j\rho_{ds} - \rho_{ds}\tilde{L}_j^+\tilde{L}_j\right\}. \quad (10)$$

The stationary density matrix can also be expressed in terms of its diagonal representation as before: $\rho_{ds} = \sum_a |a\rangle p_s(a)\langle a|$, $\{|a\rangle\}$ an orthonormal set for asymptotic, stationary time limit. These are in general different from those defined before at any given instant, t. These states now appear in defining the detailed balance condition in eq.(9) above.

We may now compare the general solution with the asymptotic solution by considering the Kullback – Leibler (KL) relative entropy, $K(\rho|\rho_s) \equiv Tr\rho\left(\ln\rho - \ln\rho_s\right) \geq 0$. The time derivative of this expression after some manipulations (after making the assumption of "microscopic detailed balance" , eq.(9) [10, 11]) is given by

$$\partial K(\rho|\rho_s)/\partial t = -\frac{1}{2}\sum_{a,b} K_{ab} p_{bs}\left[\left(\frac{p_b(t)}{p_{bs}}\right) - \left(\frac{p_a(t)}{p_{as}}\right)\right]\left[\ln\left(\frac{p_b(t)}{p_{bs}}\right) - \ln\left(\frac{p_a(t)}{p_{as}}\right)\right]$$
(11)

We obtain $\partial K(\rho|\rho_s)/\partial t \leq 0$ because $(x-y)\ln(x/y) \geq 0$ when x, y are both positive. Thus the general solution approaches the asymptotic one if the microscopic detailed balance holds.

The mean value of any physical quantity, for example the system energy, is defined by eq.(1) is expressed in the form $U = \sum_\alpha \langle\alpha|H|\alpha\rangle p(\alpha)$. If we employ the maximum entropy principle subject to this given mean energy, then the resulting density matrix is given by

$$\rho_{eq} = \exp-(\beta H)/Z, \quad Z = Tr\exp-\beta H$$
(12)

where $\beta$ is a Lagrange multiplier, the inverse temperature in units where the Boltzmann constant is taken to be unity. Z here is the partition function. In this theory then, the equilibrium density matrix given in eq.(11) leads to the known thermodynamic relation between the free energy, F, the entropy, S, and the mean energy, U: $F = -\beta^{-1}\ln Z = U - \beta^{-1}S, \quad S \equiv S_1[\rho_{eq}]$.

A quasi-equilibrium variation of this average value can come about by variations in H and in $\rho$, $dU = Tr(d\rho)H + Tr\rho(dH)$. Following [1], we may identify "work", $dW = Tr\rho(dH)$, and "quantity of heat", $dQ = Tr(d\rho)H$, quite generally. If we consider the variation to come about by time evolution, then we find that

$$dS_1/dt - \beta dQ/dt = -\sum_\alpha \frac{dp(\alpha)}{dt}(\ln p(\alpha) - \ln p_{eq}(\alpha)) = -\frac{dK(\rho|\rho_{eq})}{dt}$$
(13)

where we used the KL relation between the density matrix and its equilibrium counterpart, eq.(11), and the result mentioned above. The last term in this equation is often called the "entropy production" and this is positive. This type of thermodynamic result was deduced in [10] for weak coupling of the system with the heat bath.

## 3. TSALLIS APPROACH

This is a reworking of the results of Abe [7] in terms of the q-enropy without invoking the non-additive property or any other relation arising from zero-th law of thermodynamics as was done in [7]. Such an alternate version is illuminating in that it gives a different perspective to these results starting from statistical mechanical considerations alone. The q-entropy is given by

$$S_q = (Tr\rho^q - 1)/(1-q) \qquad (14)$$

q here is a real parameter. For q=1 we recover the BG entropy, eq.(2). If we call $c_q \equiv Tr\rho^q$, then we have the relation

$$c_q \equiv Tr\rho^q = [1 + (1-q)S_q]. \qquad (15)$$

The normalized density matrix that maximizes the q-entropy in eq.(14) subject to the constraint

$$U_q = TrP_q H, \quad P_q = \rho^q / Tr\rho^q, \quad \text{and} \quad Tr\rho = 1, \qquad (16)$$

is found to be

$$\rho_q = \left[1 - (1-q)\beta c_q^{-1}(H - U_q)\right]^{1/(1-q)} / Z_q$$

$$Z_q = Tr\left[1 - (1-q)\beta c_q^{-1}(H - U_q)\right]^{1/(1-q)} \qquad (17)$$

where $\beta$ is a Lagrange multiplier associated with the q-energy constraint, eq.(16). Henceforth we introduce $\beta_q = \beta/c_q$. From eq.(17) and (15), we also have the relationship

$$c_q = (Z_q)^{1-q} \text{ as also } Z_q = Tr\left[1 - (1-q)\beta_q(H - U_q)\right]^{q/(1-q)}. \qquad (18)$$

Calculating the derivative of the partition function, $Z_q$, with respect to $U_q$, we have

$$\frac{\partial Z_q}{\partial U_q} = \beta_q Z_q, \tag{19}$$

and so the corresponding derivative of the q-entropy however is found to be

$$\frac{\partial S_q}{\partial U_q} = \frac{1}{(1-q)} \frac{\partial}{\partial U_q}\left(Z_q^{1-q} - 1\right) = \beta. \tag{20}$$

This result is a consequence of the variational character of the relations used here and not invoking the zero-th law of thermodynamics.

Abe[7] made use of the definition of "pressure" in his derivation but here we base our derivation on only the equilibrium statistical mechanical definition of normalized q-mean energy, eq. (16) and the definitions given by eqs.(14 – 20). Thus we have the following relations:

$$Tr\rho = 1 \Rightarrow Tr(\delta\rho) = 0. \tag{21}$$

$$\delta S_q = \frac{q Tr\rho^{q-1}(\delta\rho)}{(1-q)}. \tag{22}$$

Using eq.(17) for the density matrix that maximizes the q-entropy subject to the q-mean energy, eq.(16), we obtain

$$\delta S_q = \frac{q c_q Tr\{(\delta\rho)/[1-(1-q)\beta_q(H-U_q)]\}}{(1-q)} \tag{23}$$

We now compute the change in the internal energy defined by eq.(16). This involves a change in the Hamiltonian and a change in the density matrix and this is expressed in the form:

$$\left[\delta U_q - \frac{Tr\rho^q(\delta H)}{c_q}\right] = q\frac{Tr\rho^{q-1}(\delta\rho)(H-U_q)}{c_q}, \tag{24}$$

and again using eq.(17), after a little manipulation we obtain,

$$\left[\delta U_q - \frac{Tr\rho^q(\delta H)}{c_q}\right] = \frac{qTr\{(\delta\rho)/[1-(1-q)\beta_q(H-U_q)]\}}{(1-q)\beta_q}. \tag{25}$$

This derivation is based on a perturbation theory which was not required in the traditional theory. This point will be commented upon later. Combining eqs.(23, 24), we obtain the following thermodynamic relation obtained by Abe [7], if we make the identification, $\delta S_q = \beta \delta Q_q = \beta_q c_q \delta Q_q$,

$$\left[\delta U_q - \frac{Tr\rho^q(\delta H)}{c_q}\right] = \beta_q c_q \delta Q_q. \tag{26}$$

Note that the idea of pressure is not introduced in obtaining this expression for the notion of "heat" from statistical mechanics!

Quite generally, we may identify eq.(24) as the "q - quantity of heat", $\delta Q_q$, so that combined with eq.(22), we obtain the q-version of eq.(13):

$$\delta S_q - \beta \delta Q_q = -c_q \delta_\rho K_q(\rho|\rho_q), \tag{27}$$

or equivalently,

$$(c_q)^{-1}\delta S_q - \beta_q \delta Q_q = -\delta_\rho K_q(\rho|\rho_q). \tag{28}$$

In the above the q-KL is defined as $K_q(\rho|\sigma) = \frac{1}{1-q}\{1 - Tr\rho^q \sigma^{1-q}\}$ and $\sigma = \rho_q$ is given by eq.(17). The RHS of eq.(28) may be identified with the "q-entropy production" in analogy with its q=1 counterpart discussed earlier, eq.(13).

It may be worthwhile to point out that the above derivation is not rigorous because it is based on a perturbation theory around $\rho_q$. In the traditional quantum version of the theory given in sec.2, the density matrix appeared linearly and so such a perturbation argument was not essential. But in the quantum version of the T theory, since general power of the density matrix appears, a rigorous discussion based on infinitesimal change in the density matrix is subtle and does not proceed as smoothly as in the q=1 case. What is here presented is a heuristic approach which clearly exhibits the troubles with proper quantum theory in the T formulation of quasi-equilibrium theory. With this caveat, we have here given a derivation that may be of some interest to the reader.

## 4. SUMMARY AND CONCLUDING REMARKS

We have given here the Tsallis generalization of the known (q=1) statistical mechanics derivation of the thermodynamic relations concerning entropy, quantity of heat, etc., but with the caveat mentioned above. This development is complementary to that of Abe [7] without invoking thermodynamic concepts. Here we use only statistical mechanical concepts and identify the thermodynamic entities in a parallel fashion to those given in text books [1] as in BG statistical mechanics. By analogy with the usual thermodynamics, these concepts can be used to define q-adiabatic and q-isothermal processes. Such a discussion of the corresponding q-KL version of [9] in relation to other thermodynamic relations such as Carnot cycles etc. needs a more rigorous examination and will be reported in a separate communication.

**ACKNOWLEDGEMENTS**
It is a pleasure to thank Professor Sumiyoshi Abe for encouraging me to make known this work to others besides him! Thanks are due to him for emphasizing to me the subtle nature of the quantum theory of the density matrix formulation of the T theory. Partial support of the Office of Naval Research is gratefully acknowledged.